\documentclass[english]{article}
\usepackage[T1]{fontenc}
\usepackage[latin9]{inputenc}
\usepackage{geometry}
\usepackage{amsmath}
\usepackage{amssymb}
\usepackage{graphicx}

\makeatletter
\date{}

\makeatother

\usepackage{babel}
\begin{document}
\title{\textbf{Parameter Space of Morse Oscillator}}
\author{\vspace{5mm}
\textbf{M.Y. Tan}\textsf{}\thanks{\textsf{gs58392@upm.edu.my}} $^{b}$,
\textbf{M.S. Nurisya}\thanks{\textsf{risya@upm.edu.my}} $^{a,b}$
and \textbf{H. Zainuddin}$^{a,b}$}
\maketitle
\begin{center}
{$^{a}$\em Laboratory of Computational Sciences and Mathematical
Physics, }\\
{\em Institute for Mathematical Research (INSPEM), Universiti Putra
Malaysia,}\\
{\em 43400 UPM Serdang, Selangor, Malaysia.}
\par\end{center}

\begin{center}
{$^{b}$\em Department of Physics, Faculty of Science, Universiti
Putra Malaysia, }\\
 {\em 43400 UPM Serdang, Selangor, Malaysia.}
\par\end{center}

\begin{center}
\par\end{center}

\begin{center}
\vspace{0.5mm}
\par\end{center}
\begin{abstract}
We present the analysis of mathematical structure of SU(2) group,
specifically the commutation relation between raising and lowering
operators of the Morse oscillator. The relationship between the commutator
of operators and other parameters of Morse oscillator is investigated.
We show that the mathematical structure of operators which depends
on the parameters of Morse oscillator may change our conventional
expectation. The parameter space of Morse oscillator is visualized
to scrutinize the mathematical relations that are related to the Morse
oscillator. This parameter space is the space of possible parameter
values that depend on the depth of the Morse potential well and other
parameters. The algorithm that we present is also applicable to other
quantum systems with certain modifications.\\
\\
\textbf{Keywords:} Morse potential, ladder operators, commutation
relation, eigenvalue, parameter space.
\end{abstract}

\section{Introduction}

In the early development of quantum mechanics, the exactly solvable
potentials attracted researchers' attention. Different exactly solvable
potentials were introduced such as Coulomb, Morse, Rosen-Morse, P\"{o}schl-Teller
and Eckart potentials. There are various studies that can be conducted
regarding these potentials. For instance, the application of factorization
method \cite{key-1a,key-2b} connected to raising and lowering operator
method, supersymmetry and shape invariance \cite{key-2b,key-3c,key-4d}.
Furthermore, the concept of ladder operators had been further extended
to include its associated Lie algebras \cite{key-5e}.

The Morse potential is important in the calculations in molecular
spectroscopy and diatomic molecule's modelling. It is often investigated
analytically. A unified description of the position-space wave functions,
the momentum-space wave functions, and the phase-space Wigner functions
for the bound states of a Morse oscillator was presented \cite{key-6f}.
The constructions of its ladder operators and its algebraic structure
in terms of SU(2) were made in the previous studies \cite{key-7g}.

The objective of this study is to analyse the commutation relation
between the ladder operators for Morse potential. We show the relationship
between the parameters in the ladder operators and commutator of ladder
operators for the Morse oscillator. Besides, we also generate the
plots of parameter space of Morse oscillator to examine some mathematical
relations. This parameter space is the space of possible integer parameter
values that depend on the depth of the Morse potential well and other
parameters implicitly such as the mass of molecule and the ``width''
of the potential. This work is organized as follows. In the following
section, we provide the Morse potential and its solutions. We also
introduce the lowering and raising operators of Morse oscillator that
are constructed directly from its eigen-wave function. In Section
3, we analyse the commutation relation between the ladder operators.
We also establish different mathematical forms of an operator deduced
from the commutation relation. In Section 4, we present the implementation
of an algorithm to investigate the operator that is established through
the commutator of ladder operators. We display the plots of the parameter
space of Morse potential in Section 5. We later discuss the equality
of eigenvalues obtained through two different perspectives. These
have not much been discussed in the literature previously. Our analysis
expands the understanding of commutation relation for Morse quantum
system. In the final section, we present our conclusions.

\section{Ladder Operators for Morse Potential}

The Morse potential has the following form \cite{key-8h}:

\begin{equation}
V(x)=V_{0}(e^{-2\beta x}-2e^{-\beta x}),\label{eq:1}
\end{equation}
where $V_{0}>0$ corresponds to its depth, $x$ is the relative displacement
from the equilibrium position of the atoms and $\beta$ is related
to the ``width'' of the potential. The associated Schr\"{o}dinger
equation with different mathematical form of Eq. (\ref{eq:1}) is
then given by \cite{key-6f}, 

\begin{equation}
-\frac{\hbar^{2}}{2m}\frac{d^{2}\psi}{dx^{2}}+V_{0}(1-e^{-\beta x})^{2}\psi=E\psi,\label{eq:2}
\end{equation}
where $m$ is the mass of the molecule. The solutions of Eq. (\ref{eq:2})
have the form \cite{key-6f}

\begin{equation}
\psi_{n}^{v}(y)=N_{n}^{v}e^{-\frac{y}{2}}y^{s}L_{n}^{2s}(y),\label{eq:3}
\end{equation}
where $L_{n}^{2s}(y)$ are the associated Laguerre functions. There
is a coordinate transformation of argument $x$ in which $y=ve^{-\beta x}.$
The normalization constant $N_{n}^{v}$ is

\begin{equation}
N_{n}^{v}=\sqrt{\frac{\beta(v-2n-1)\Gamma(n+1)}{\Gamma(v-n)},}\label{eq:4}
\end{equation}
where the parameters $v$ and $s$ are related to the depth of the
potential well and energy eigenvalue respectively. The parameters
$v$ and $s$ are 

\begin{equation}
v=\sqrt{\frac{8mV_{0}}{\beta^{2}\hbar^{2}}},\qquad s=\sqrt{\frac{-2mE}{\beta^{2}\hbar^{2}}},\label{eq:5}
\end{equation}
with the constraint condition on solutions of Eq. (\ref{eq:2}) such
that

\begin{equation}
2s=v-2n-1.\label{eq:6}
\end{equation}
The annihilation and creation operators for the Morse wave functions
have been established with the relations related to the associated
Laguerre polynomials \cite{key-7g}. The annihilation operator for
the Morse oscillator has the form

\begin{equation}
\hat{K}_{-}=-\left[\frac{d}{dy}(2s+1)-\frac{1}{y}s(2s+1)+\frac{v}{2}\right]\sqrt{\frac{s+1}{s}},\label{eq:7}
\end{equation}
which it obeys the following equation:

\begin{equation}
\hat{K}_{-}\psi_{n}^{v}(y)=k_{-}\psi_{n-1}^{v}(y),\label{eq:8}
\end{equation}
where 
\begin{equation}
k_{-}=\sqrt{n(v-n)}.\label{eq:9}
\end{equation}
The creation operator is defined by

\begin{equation}
\hat{K}_{+}=\left[\frac{d}{dy}(2s-1)+\frac{1}{y}s(2s-1)-\frac{v}{2}\right]\sqrt{\frac{s-1}{s}},\label{eq:10}
\end{equation}
in which it satisfies the equation

\begin{equation}
\hat{K}_{+}\psi_{n}^{v}(y)=k_{+}\psi_{n+1}^{v}(y),\label{eq:11}
\end{equation}
where 
\begin{equation}
k_{+}=\sqrt{(n+1)(v-n-1).}\label{eq:12}
\end{equation}

One can construct the commutator $\left[\hat{K}_{+},\hat{K}_{-}\right]$
from the relations (\ref{eq:8}) and (\ref{eq:11}) that acts on the
wave function of Morse oscillator

\begin{equation}
\left[\hat{K}_{+},\hat{K}_{-}\right]\psi_{n}^{v}(y)=2k_{0}\psi_{n}^{v}(y),\label{eq:13}
\end{equation}
where the eigenvalue is referred to as $k_{0}^{'}=2k_{0}$ with

\begin{equation}
k_{0}=n-\frac{v-1}{2}.\label{eq:14}
\end{equation}
Thus, we can define the operator $\hat{K}_{0}$ using Eq. (\ref{eq:14})
as

\begin{equation}
\hat{K}_{0}=\hat{n}-\frac{v-1}{2}.\label{eq:15}
\end{equation}
It can also be rewritten in terms of differential operator with the
help of the associated Schr\"{o}dinger equation \cite{key-7g}

\begin{equation}
\left(y\frac{d^{2}}{dy^{2}}+\frac{d}{dy}-\frac{s^{2}}{y}-\frac{y}{4}+\frac{v}{2}\right)\psi_{n}^{v}(y)=0,\label{eq:16}
\end{equation}
from which this operator $\hat{K}_{0}$ is established as

\begin{equation}
\hat{K}_{0}=\left(y\frac{d^{2}}{dy^{2}}+\frac{d}{dy}-\frac{s^{2}}{y}-\frac{y}{4}+n+\frac{1}{2}\right).\label{eq:17}
\end{equation}
The operators $\hat{K}_{\pm}$ and $\hat{K}_{0}$ satisfy the following
commutation relation

\begin{equation}
\left[\hat{K}_{+},\hat{K}_{-}\right]=2\hat{K}_{0}=\hat{K}_{0}^{'}.\label{eq:18}
\end{equation}
It is important to point out that the operators $\hat{K}_{\pm}$ and
$\hat{K}_{0}$ are dependent on two parameters, namely any combination
of two elements from the set of $\{s,n,v\}$. Equation (\ref{eq:18})
has been realized that it satisfies the algebra of SU(2).

\section{Commutation relation between ladder operators $\hat{K}_{+}$ and
$\hat{K}_{-}$}

There are two perspectives that we can look at from the commutation relation
between ladder operators (\ref{eq:18}). The first perspective is
that we consider the eigenstates of the Schr\"{o}dinger equation
that is given by Eq. (\ref{eq:3}) and the relations (\ref{eq:8})
and (\ref{eq:11}) of the ladder operators. This is exactly what Eq.
(\ref{eq:13}) implies. The second perspective is that we consider
any arbitrary test function and the differential form of the ladder
operators. In the case without the consideration of the changes in
parameter involved, we have

\begin{align}
\left[\hat{K}_{+},\hat{K}_{-}\right]f & =\left[\left(\frac{d}{dy}(2s-1)+\frac{1}{y}s(2s-1)-\frac{v}{2}\right)\sqrt{\frac{s-1}{s}},\right.\nonumber \\
 & \quad\left.-\left(\frac{d}{dy}(2s+1)-\frac{1}{y}s(2s+1)+\frac{v}{2}\right)\sqrt{\frac{s+1}{s}}\right]f\nonumber \\
 & =2\sqrt{1-\frac{1}{s^{2}}}s(2s-1)(2s+1)\left[\frac{d}{dy},\frac{1}{y}\right]f\nonumber \\
 & =-\frac{2\sqrt{1-\frac{1}{s^{2}}}s(2s-1)(2s+1)}{y^{2}}\cdot f\nonumber \\
 & =\frac{2\left(s^{2}-1\right)^{\frac{1}{2}}\left(1-4s^{2}\right)}{y^{2}}\cdot f\nonumber \\
 & =2\hat{K}_{0}\cdot f\nonumber \\
 & =\hat{K}_{0}^{'}\cdot f.\label{eq:19}
\end{align}
We have derived an operator $\hat{K}_{0}^{'}$. There is an additional
constraint that should be placed on this derived operator. The constraint
that we want to look at is that the derived operator $\hat{K}_{0}^{'}$
satisfies the eigenvalue equation in the form

\begin{equation}
\hat{K}_{0}^{'}\psi_{n}^{v}(y)=k_{0}^{'}\psi_{n}^{v}(y).\label{eq:20}
\end{equation}
Since this derived operator does not satisfy the eigenvalue equation,
we consider another possible condition. In the case with the consideration
of the changes in parameter involved, we have

\begin{align}
\left[\hat{K}_{+},\hat{K}_{-}\right]f & =\left[\hat{K}_{+}^{n-1,v}\hat{K}_{-}^{n,v}-\hat{K}_{-}^{n+1,v}\hat{K}_{+}^{n,v}\right]f\nonumber \\
 & =\left[\hat{K}_{+}^{s+1,v}\hat{K}_{-}^{s,v}-\hat{K}_{-}^{s-1,v}\hat{K}_{+}^{s,v}\right]f\nonumber \\
 & =s\left(-\frac{8}{y}\frac{d}{dy}-8\frac{d^{2}}{dy^{2}}+\frac{8s^{2}}{y^{2}}-\frac{4v}{y}\right)f\nonumber \\
 & =\hat{K}_{0}^{'}\cdot f.\label{eq:21}
\end{align}
This derived operator from Eqs. (\ref{eq:21}) is examined through
an algorithm by computing its eigenvalues. One should keep in mind
that there are three definitions for operator $\hat{K}_{0}^{'}$ or
$\hat{K}_{0}$, as shown in Eqs. (\ref{eq:15}), (\ref{eq:17}) and
(\ref{eq:21}). Hence, the eigenvalues $k_{0}^{'}$ can be obtained
in three different ways. However, the eigenvalues $k_{0}^{'}$ calculated
using the definitions in Eqs. (\ref{eq:15}) and (\ref{eq:17}) are
always equal due to their mathematical constructions.

\section{Implementation in Mathematica}

\begin{figure}
\centering{}\includegraphics[scale=0.5]{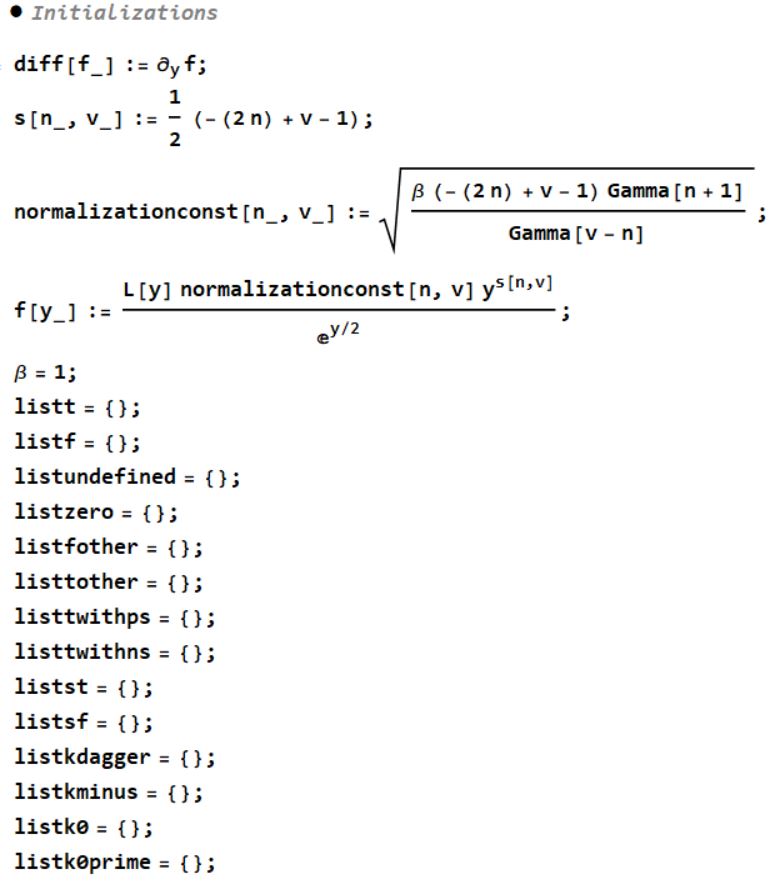}\caption{The first block of Mathematica code to implement an algorithm to compute
and compare the eigenvalues in the two different perspectives.\label{fig:1}}
\end{figure}

The Mathematica code is divided into six different blocks for an implementation
of the algorithm to compute and compare the eigenvalues $k_{0}^{'}$
in the two different perspectives. Figure \ref{fig:1} shows the first
block of Mathematica code of an algorithm. In the first block, we
define the differential operator $\frac{d}{dy}$ to construct the
operators $\hat{K}_{\pm}$, $\hat{K}_{0}$ and $\hat{K}_{0}^{'}$
later. Besides, we define the constraint condition, normalization
constant and the wave function of Morse potential according to Eqs.
(\ref{eq:6}), (\ref{eq:4}) and (\ref{eq:3}) respectively. The value
of the parameter $\beta$ is set to $1$, in which its value will
not affect the result of our analysis. Lastly, we define some empty
sets for graphics later.

\begin{figure}
\centering{}\includegraphics[scale=0.54]{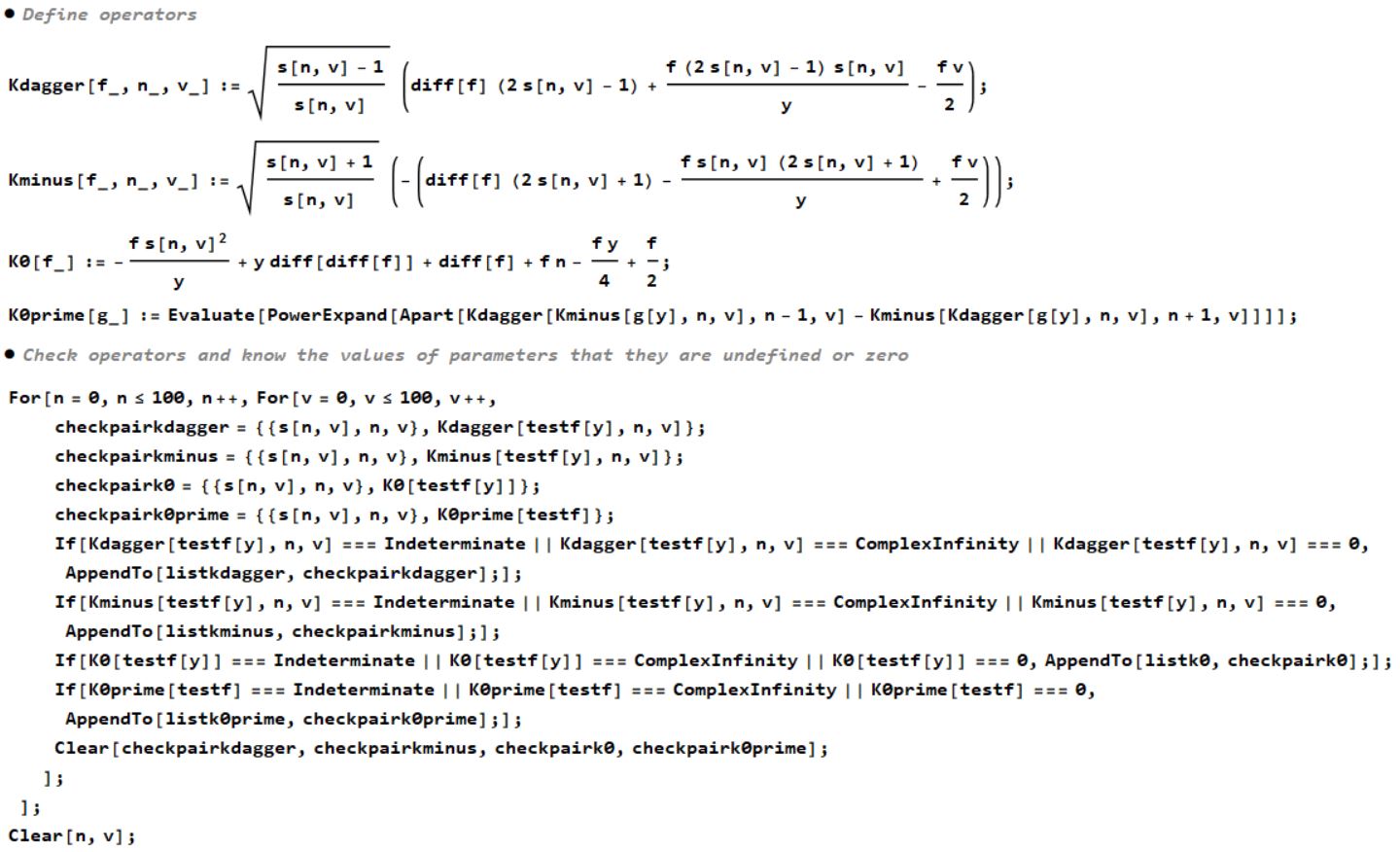}\caption{The second and third blocks of Mathematica code to implement an algorithm
to compute and compare the eigenvalues in the two different perspectives.\label{fig:2}}
\end{figure}

After executing the first block, we continue to execute the second
and third blocks of the code as shown in figure \ref{fig:2}. Second
block defines the operators $\hat{K}_{\pm}$, $\hat{K}_{0}$ and $\hat{K}_{0}^{'}$
in terms of differential operator. The next block is to determine
the parameter values which lead to undefined or zero operators. 

\begin{figure}
\centering{}\includegraphics[scale=0.47]{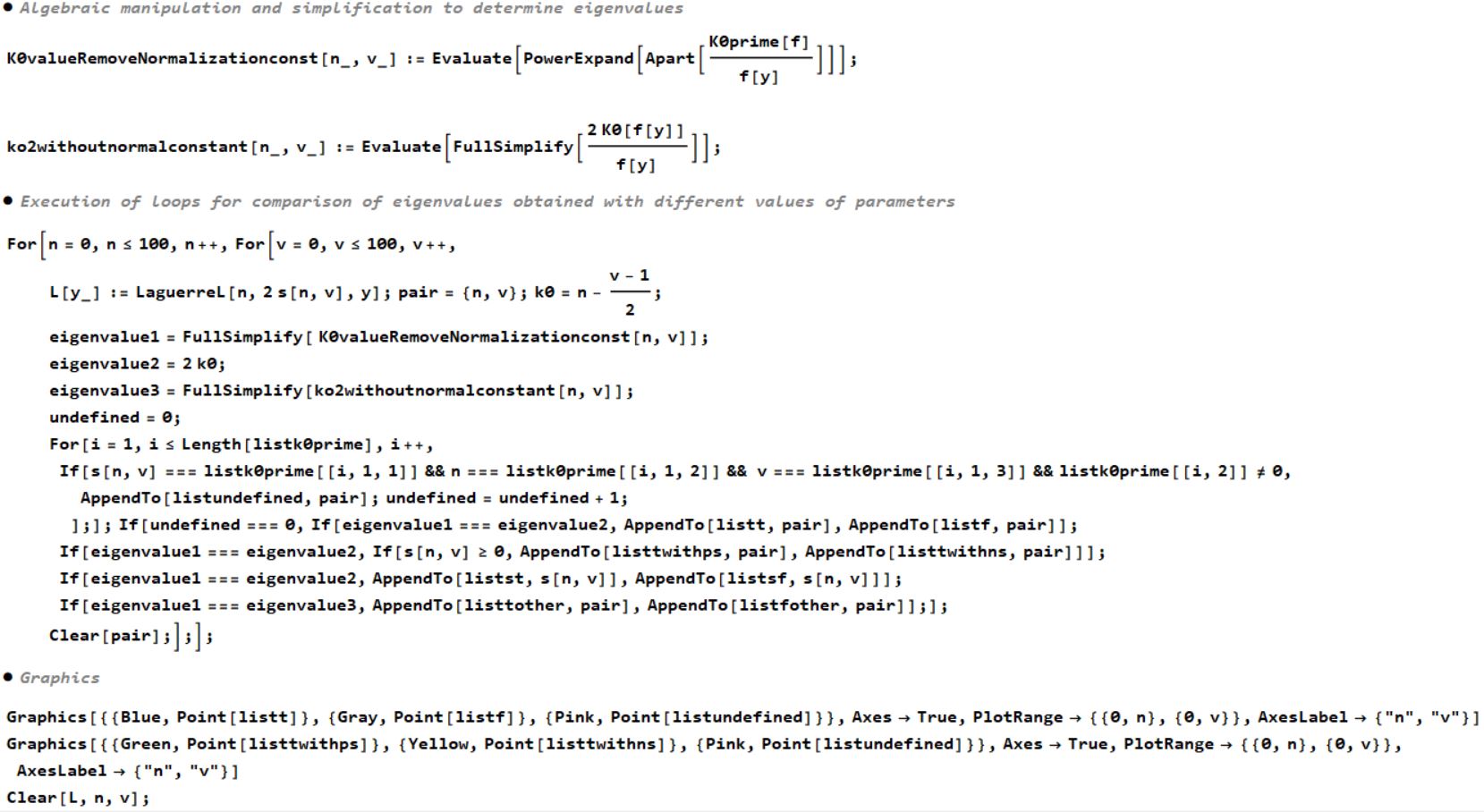}\caption{The fourth, fifth and sixth blocks of Mathematica code to implement
an algorithm to compute and compare the eigenvalues in the two different
perspectives.\label{fig:3}}
\end{figure}

The fourth block of code defines the functions to calculate the eigenvalues.
The first function is constructed with the algebraic manipulation
of Eq. (\ref{eq:20}). The second function in this block evaluates
the eigenvalues, $k_{0}^{'}$ through the operator $\hat{K}_{0}$
as given by Eq. (\ref{eq:17}). The ``PowerExpand'', ``Apart''
and ``FullSimplify'' commands simplify the calculation, including
the cancellation of normalization constant. The execution of loops
takes place in the fifth block of code as depicted in figure \ref{fig:3}.
The integer parameters $n$ and $v$ are set to be in the range from
$0$ to $100.$ In the loop body, we define the associated Laguerre
polynomial with the Mathematica command ``LaguerreL''. We obtain
the eigenvalues for each pair of parameters $\left(n,v\right)$ and
denote them in the Mathematica code by ``eigenvalue1'', ``eigenvalue2''
and ``eigenvalue3''. Since the ``eigenvalue2'' and ``eigenvalue3''
are always equal, hence the comparison between the ``eigenvalue1''
and ``eigenvalue2'' becomes our interest. There are a lot of if
statements in the loop for the extractions of values into the sets
for the visualization purpose. The sixth block as shown in figure
\ref{fig:3} creates our visualization scheme. Two plots on the parameter
space of $n$ and $v$ of the Morse oscillator will be displayed. 

\section{Results and Discussion}

\begin{figure}
\centering{}\includegraphics[scale=0.5]{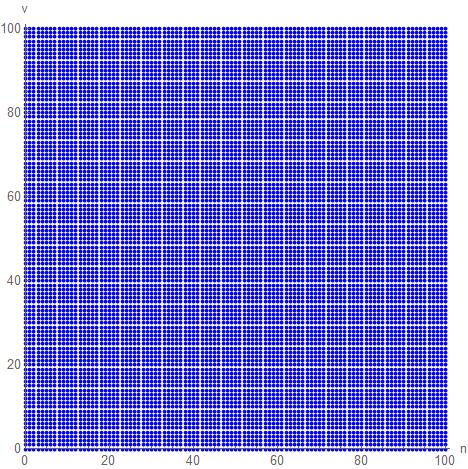}\caption{(color online) The parameter space of $n$ and $v$ of the Morse oscillator.
Blue dot represents the equality of eigenvalue.\label{fig:4}}
\end{figure}

With the algorithm, we calculate the eigenvalues with different mathematical
formulas and then compare them. We want to verify that the derived
operator $\hat{K}_{0}^{'}$ satisfies the eigenvalue equation (\ref{eq:20}).
The parameter space of $n$ and $v$ of the Morse oscillator is plotted
to visualise our analysis as shown in figure \ref{fig:4}. We check
the equality of eigenvalues $10201$ times in total for pairs of $(n,v)$.
Since we do not determine the mathematical expression of the eigenvalues
$k_{0}^{'}$ of the derived operator $\hat{K}_{0}^{'}$ explicitly,
we can only claim that we are highly confident that the derived operator
$\hat{K}_{0}^{'}$ satisfies the eigenvalue equation (\ref{eq:20}).
Two perspectives on the commutation relation are consistent, because
the equality of eigenvalues obtained through two different perspectives
seems to hold. However, the conclusion is not definite. There is a
caveat which concerns the computation. The derived operator $\hat{K}_{0}^{'}$
is zero at parameter $s$ equals $0$, which is ambiguous. Coincidentally,
we compute its eigenvalues to be also zero at $s=0$, which is a trivial
solution.

\begin{figure}
\centering{}\includegraphics[scale=0.5]{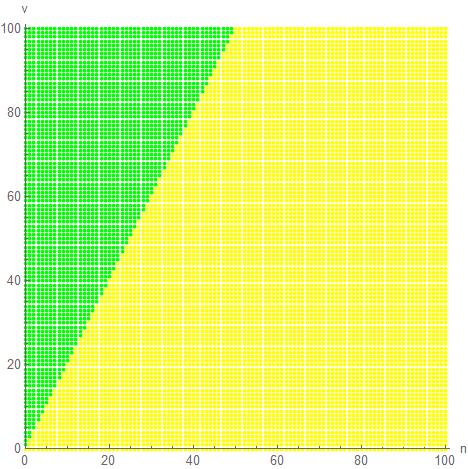}\caption{(color online) The parameter space of $n$ and $v$ of the Morse oscillator.
Green dot represents the eigenvalue with the parameter $s$ that is
greater than or equal to zero while yellow dot represents the eigenvalue
with the parameter $s$ that is less than zero.\label{fig:5}}
\end{figure}

Figure \ref{fig:5} shows the plot of parameter space of $n$ and
$v$ with green, and yellow dots. According to the definition of the
parameter $s$ in Eqs. (\ref{eq:5}), the values of the parameter
$s$ can only take non-negative real numbers, $\mathbb{R}_{\geq0}=\left\{ x\in\mathbb{R}\mid x\geq0\right\} $.
Besides that, the parameter $s$ is further constrained by Eq. (\ref{eq:6})
which it leads to the inequality $v\geq2n+1$. This is clearly indicated
in the plot, see figure \ref{fig:5}.

\section{Conclusions}

The present paper provides, for the first time, a detailed analysis
on a particular commutation relation of the dynamic group SU(2) for
the Morse oscillator. There is no evidence of detailed discussion
on a specific commutation relation between the generators of the dynamic
group SU(2) for the Morse potential before. The question may be raised
regarding the interpretation of commutation relation between operators
that depend on some parameters. The consideration for some changes
in the parameter of the operators in a commutation relation is needed.
We speculate that the derived operator may be related to the adjoint
of operator $\hat{K}_{0}$ in Eq. (\ref{eq:17}). This may be the
reason that this derived operator satisfies the eigenvalue equation
(\ref{eq:20}). An operator can be undefined or zero at some parameter
values if it relies on the parameters explicitly. The mathematical
structure of the differential form of operators and commutator that
depends on parameters may restrict the region in the parameter space.
The regions in the parameter space can be bounded by different conditions.
For example, the parameter $s$ forms a boundary in the parameter
space at zero due to the inequality $s\geq0$ imposed by its definition.
We have paid attention to the interpretation of commutation relation
that involves operators with parameters. We choose the Morse oscillator
to illustrate our point of view. The algorithm that is written in
Mathematica code may also be applied to other quantum systems after
some modifications. The modern-day computing power allows us to scrutinize
the mathematical relations with the help of computer. We believe that
our analysis will provide better understanding of the loophole that
may occur in a mathematical relation, for instance, the commutation
relation between the ladder operators for the Morse potential.

\end{document}